\documentclass[conference]{IEEEtran}
\IEEEoverridecommandlockouts 

\usepackage[cmex10]{amsmath}  
\usepackage{amssymb}
\usepackage{graphicx,color}   

\usepackage{ifxetex}
\ifxetex
\usepackage{xunicode}
\else
\usepackage[T1]{fontenc}
\usepackage[utf8]{inputenc}
\usepackage{microtype}
\fi

\graphicspath{{../pictures/}}

\usepackage{cite}  

\newtheorem{theorem}{Theorem}

\newtheorem{lemma}{Lemma}

\newtheorem{proposition}{Proposition}

\newcommand{\mais}{\textsf{MAIS} }
\newcommand{\minrank}{\textsf{minrank} }

\newcommand{\figgap}{\vspace{-1.4ex}}

\IEEEoverridecommandlockouts
\begin{document}

\title{Linear Codes are Optimal for Index-Coding Instances with Five or Fewer Receivers}
\author{\vspace{-1ex}\IEEEauthorblockN{Lawrence Ong}\vspace{-2ex}
\thanks{Lawrence Ong is the recipient of an Australian Research Council Discovery Early Career Researcher Award  (project number DE120100246).}
}
\maketitle

\begin{abstract}
We study zero-error unicast index-coding instances, where each receiver must perfectly decode its requested message set, and the message sets requested by any two receivers do not overlap. We show that for all these instances with up to five receivers, linear index codes are optimal. Although this class contains 9847 non-isomorphic instances, by using our recent results and by properly categorizing the instances based on their graphical representations, we need to consider only 13 non-trivial instances to solve the entire class.
This work complements the result by Arbabjolfaei et al.\ (ISIT 2013), who derived the capacity region of all unicast index-coding problems with up to five receivers in the \textit{diminishing-error} setup. They employed random-coding arguments, which require infinitely-long messages. We consider the \textit{zero-error} setup; our approach uses graph theory and combinatorics, and does not require long messages.

\end{abstract}

\section{Introduction}

We study index-coding problems, where a sender broadcasts different messages to multiple receivers, each knowing of some messages a priori. This is an open problem, and only a few classes of the problems have been solved. Based on our recent results~\cite{ong13iccsubmitted}, we solve the problems for up to five receivers and show that linear index codes are optimal for all these problems. 

\subsection{Index Coding}

An $n$-receiver index-coding instance is defined as follows: A sender has a binary message set $\mathcal{X} \triangleq \{x_1, x_2, \dotsc, x_m\}$, where each $x_i \in \{0,1\}$. It maps these messages to a length-$\ell$ binary codeword, denoted by $\epsilon(\mathcal{X})$, and broadcasts this codeword to $n$ receivers. Each receiver $i \in \{1,2,\dotsc, n\}$ knows some messages a priori, $\mathcal{K}_i \subset \mathcal{X}$, and wants some messages from the sender, $\mathcal{W}_i \subseteq \mathcal{X}$.

 We say that the encoding map $\epsilon(\mathcal{X})$ is an \textit{index code} if and only if there exists a decoding function $\delta_i$ for each receiver $i$ such that the receiver can decode (without error) the message set it wants from the codeword sent by the sender and the message set it knows a priori, i.e., $\delta_i(\epsilon(\mathcal{X}),\mathcal{K}_i) = \mathcal{W}_i$. The aim is to find the shortest index codelength, denoted by $\ell^*$.

\subsection{Unicast Index Coding and Side-Information Graph}

In this paper, we consider the \textit{unicast} index-coding instances, where no two receivers want a same message, i.e., $\mathcal{W}_i \cap \mathcal{W}_j = \emptyset$ for all $i \neq j$. We assume that each receiver requests only one message, i.e., $|\mathcal{W}_i| = 1$ for all $i$. This assumption is of no loss of generality, as a receiver who requests, say, two messages is equivalent (as far as index codes are concerned) to two receivers, both knowing the same prior messages, and each requesting a different message. 

We can remove any receiver who does not request any message, and any message not requested by any receiver. So, we have $n$ receivers and $n$ messages, and we denote by $x_i$ the message requested by receiver $i$. 

\begin{figure}[t]
\centering
\includegraphics[width=0.8cm]{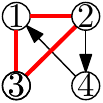}
\caption{The side-information graph for a four-receiver index-coding instance where receiver 1 knows $(x_2,x_3)$, receiver 2 knows $(x_1,x_3,x_4)$, receiver 3 knows $(x_1,x_2)$, receiver 4 knows $x_1$. An edge (drawn with a red thick line) represents arcs in both directions. Here $1-2-3-1$ forms an undirected cycle, and $1 \rightarrow 2 \rightarrow 4 \rightarrow 1$ forms a directed cycle (or simply, cycle).}
\label{fig:example}
\figgap
\end{figure}

These index-coding instances can be completely described using directed graphs, known as \textit{side-information} graphs~\cite{baryossefbirk11}. A side-information graph consist of $n$ vertices (where $n$ is the number of receivers), and an arc from vertex $i$ to vertex $j$ if and only if receiver $i$ knows message $x_j$ a priori. We denote an arc from $i$ to $j$ by $i \rightarrow j$. Further, we represent a bidirectional arc by an edge. An edge between $i$ and $j$, denoted by $i - j$, represents both arcs $i \rightarrow j$ and $j \rightarrow i$. An undirected cycle is a cycle formed by edges, and a directed cycle is one formed by arcs. For the rest of this paper, we refer to directed cycles simply as cycles. 
See Figure~\ref{fig:example} for an example.

For a side-information graph $G$, we denote the set of vertices by $V(G)$, the order (i.e., the number of vertices) of $G$ by $|V(G)|$, and the optimal index codelength by $\ell^*(G)$. 

\subsection{Existing Results for Unicast Index Coding}

The most general (i.e., applicable to all $G$) lower and upper (achievability) bounds to $\ell^*$, for unicast index coding, are~\cite{baryossefbirk11}
\begin{equation}
\mais(G) \leq \ell^*(G) \leq \minrank(G),
\end{equation}
where $\mais(G)$ is the order of a \textit{maximum acyclic induced subgraph} (MAIS) of $G$, and  $\minrank(G)$ is a function of the graph $G$ based on it \textit{adjacency matrix}. An MAIS is an acyclic vertex-induced subgraph that has the largest number of vertices. The minrank function returns the best \textit{linear} index codelength. Both the MAIS upper bound and the minrank lower bound are NP-hard to compute~\cite{karp72,peeters96}, and both have been shown to be loose in some instances~\cite{baryossefbirk11,lubertzkystav09}. This implies that linear index codes, though having practical advantages of simplifying encoding and decoding, are not necessarily optimal.



It is possible to find $\ell^*$ by brute force. This can be done by forming the \textit{confusion graph} of $G$, and finding the chromatic number (which is NP-complete) of the confusion graph. This gives $\ell^*(G)$. This method is intractable as $|V(G)|$ grows because the order of the confusion graph is $2^{|V(G)|}$.

To date, $\ell^*$ has been characterized (without resorting to the brute-force search) for only a small number of classes of side-information graphs~\cite{baryossefbirk11}: (a) acyclic, (b) undirected\footnote{A side-information graph is undirected if and only if  it contains only edges, i.e., if receiver $i$ knows $x_j$, then receiver $j$ must also know $x_i$, for all $i$ and $j$.} and perfect, (c) undirected odd holes, or (d) undirected odd anti-holes. Recently, Yu and Neely~\cite{yuneely13} represented unicast index-coding instances using bipartite graphs, and found $\ell^*$  for all planar bipartite graphs. 
For all these classes, linear index codes have been shown to be optimal, i.e., $\ell^* = \minrank(G)$.

We have recently characterized $\ell^*$ for a new class of index-coding instances, in which an MAIS can be formed by removing two or fewer vertices from $G$, stated as follows~\cite{ong13iccsubmitted}:
\begin{lemma}\label{lem:mais-2}
If $\mais(G) \geq |V(G)| - 2$, then $\ell^*(G) = \mais(G)$, and $\ell^*(G)$ is achievable by linear index codes, which implies that $\ell^*(G) = \minrank(G)$.
\end{lemma}

In this paper, we build on Lemma~\ref{lem:mais-2} to find $\ell^*$ for any $G$ up to five vertices, and show that $\ell^*(G) = \minrank(G)$. The instance in Figure~\ref{fig:example} cannot be described using either an acyclic $G$, an undirected $G$, or a planar bipartite graph. Hence existing results~\cite{baryossefbirk11,yuneely13} do not subsume the result of this paper.

In a slightly different setup, a similar result has been obtained by Arbabjolfaei et al.~\cite{arbabjolfaei13}, who derived the optimal \textit{code rate} for all unicast index-coding instances with up to five receivers. In their setup, each message $x_i$ contains $BR_i$ bits (for $i \in \{1,\dotsc, n\}$), and the sender broadcasts a $B$-bit codeword. The aim is to find the region of all rate tuples $(R_1, \dotsc, R_n)$ such that the probability that any receiver makes a decoding error tends to zero as $B$ tends to infinity. This paper considers a different setup (to find $\ell^*$), and uses a different approach:
\begin{enumerate}
\item We consider the zero-error setup; they consider the diminishing-error setup. 
\item We consider one-bit messages; their coding scheme requires infinitely long messages of $BR_i$ bits, $B \rightarrow \infty$.
\item Our achievability uses linear codes (simple XOR of messages, which are readily implementable); their coding scheme is based on random coding arguments.
\end{enumerate}
Note that a zero-error/short-message solution is a solution to the diminishing-error/infinitely-long-message problem, but not vice versa. We leave the comparison as our future work.


\subsection{Our Contributions}

In this paper, we establish the following:

\begin{theorem} \label{theorem:main}
If $|V(G)|\leq 5$, then
\begin{equation} 
\ell^*(G) = \minrank(G). \label{eq:linear-opt}
\end{equation}
\end{theorem}

We note that it is theoretically possible to arrive at the results in Theorem~\ref{theorem:main} by brute force, i.e., (i) computing $\minrank(G)$ for all non-isomorphic directed graphs with five or fewer vertices; (ii) computing the chromatic number of all corresponding confusion graphs, which gives $\ell^*(G)$; and (iii) showing that they are equal for every directed graph.
However, this method is intractable; for graphs with five vertices alone, there are 1,048,576 directed graphs, out of which 9608 are non-isomorphic~\cite{sloane10}, and each five-vertex directed graph corresponds to a  confusion graph with 32 vertices. 

Instead, we prove Theorem~\ref{theorem:main} by classifying $G$ according to the number of undirected cycles. We then show that \eqref{eq:linear-opt} holds for each category.  The main difficulty is to correctly identifying the classification such that each contains a special \textit{arc-deleted subgraph} (i.e., a subgraph with the same vertices but \textit{possibly} fewer arcs) for which we can easily find an optimal index code (which is incidentally linear).
Using our method, we need to evaluate the chromatic number of only two confusion graphs.

Our method also serves as a potential approach to characterizing $\ell^*$ for graphs of higher orders---though not straightforward. We leave this for future investigation.

\section{Proof of Theorem~\ref{theorem:main}}

We now prove Theorem~\ref{theorem:main} by considering graphs of different orders. 
For $|V(G)|=1$, the optimal index codelength is trivial: $\ell^*(G)= \mais(G) = \minrank(G)=1$.

Note that for any graph, we must have that $1 \leq \mais(G) \leq |V(G)|$. This means for $|V(G)| =$ 2 or 3, the condition in Lemma~\ref{lem:mais-2} is always true, and hence the optimal index codelength is $\ell^*(G) = \mais(G) =  \minrank(G)$.

For $|V(G)|=4$, we again use Lemma~\ref{lem:mais-2} to show that linear codes are optimal and achieve $\ell^*(G) = \mais(G) = \minrank(G)$ if $\mais(G) = 2,3,4$. For the remaining case where $\mais(G) = 1$, 
any two-vertex induced subgraph  must contain a cycle (i.e., arcs in both directions between any two vertices); otherwise $\mais(G) \geq 2$. In other words, each receiver $i$ know all other messages $\{x_j: j \in \{1,2,3,4\} \setminus \{i\}\}$. So, sending $x_1 \oplus x_2 \oplus x_3 \oplus x_4$ satisfies all receivers' requirements, and the one-bit MAIS lower bound is achievable.

For $|V(G)|=5$, using Lemma~\ref{lem:mais-2}, linear codes are optimal and achieve $\ell^*(G) = \mais(G) = \minrank(G)$ if $\mais(G) = 3,4,5$. Also, if $\mais(G)=1$, we can use the same argument for $|V(G)|=4$ to show that one coded bit of $x_1 \oplus x_2 \oplus x_3 \oplus x_4 \oplus x_5$ is achievable and is hence optimal.

For all the above cases, linear codes are optimal, and the MAIS lower bound is tight. To complete the proof of Theorem~\ref{theorem:main}, we need to prove the remaining case (the main contribution of this paper), stated in the following proposition:
\begin{proposition} \label{proposition}
If
\begin{itemize}
\item the graph has five vertices, i.e., $|V(G)|=5$, and
\item we need to remove two (and not fewer) vertices to get an acyclic induced subgraph, i.e., $\mais(G) = 2$,
\end{itemize}
then linear codes are optimal, i.e., $\ell^*(G) = \minrank(G)$. 
\end{proposition}

We present the proof of Proposition~\ref{proposition} in Section~\ref{sec:3}. As we shall see, the MAIS lower bound is no longer  always tight.

\section{Proof of Proposition~\ref{proposition}: Linear Codes are Optimal when $|V(G)|=5$ and $\mais(G)=2$} \label{sec:3}

First, note that the condition $\mais(G)=2$ is equivalent to
\begin{itemize}
\item Any induced subgraph of $G$ with three vertices must contain a cycle. 
\end{itemize}
Otherwise, $\mais(G) \geq 3$ by considering the 3-vertex induced subgraph without a cycle.

\subsection{Four categories}

As the proof is rather involved, we propose to divide the set of graphs with $|V(G)|=5$ and $\mais(G)=2$ into four categories according to the number of undirected cycles in $G$:
\begin{enumerate}
\item There is no undirected cycle.
\item There exists an undirected cycle of length three.
\item There is no undirected cycle of length three, but there exists an undirected cycle of length four.
\item There is no undirected cycle of length three or four, but there exists an undirected cycle of length five.
\end{enumerate}

Note that, by definition, there cannot be any undirected cycle of length less than three.

\subsection{Two Useful Lemmas}

We first prove two lemmas to be used subsequently:
\begin{lemma} \label{lem:remove-add-arcs}
Let $G$ be an arc-deleted subgraph of $G^+$, and $G^-$ be an arc-deleted subgraph of $G$. Then,
\begin{equation}
\ell^*(G^+) \leq \ell^*(G) \leq \ell^*(G^-),
\end{equation}
and an index code for $G^-$ is an index code for $G$ and $G^+$.
\end{lemma}

\begin{IEEEproof}
Each receiver in $G^+$ has at least the prior messages that it has in $G$, and it requests for the same message (i.e., receiver $i$ requests for $x_i$). So, any index code for $G$ satisfies all decoding requirements for $G^+$ and hence is an index code for $G^+$. This proves $\ell^*(G^+) \leq \ell^*(G)$. By repeating the same argument, we have $\ell^*(G) \leq \ell^*(G^-)$.
\end{IEEEproof}

\begin{lemma} \label{lem:4-cycle}
If $|V(G)|=5$ and $\mais(G)=2$, then the induced subgraph of any four vertices must contain an edge. 
\end{lemma}

\begin{figure}[t]
\centering
\includegraphics[width=0.8cm]{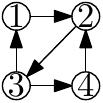}
\caption{If there is no edge in the graph, then $\{1,3,4\}$ cannot contain a cycle. }
\label{fig:4-cycle}
\figgap
\end{figure}

\begin{IEEEproof}
Suppose, for the sake of contradiction, that there is an induced subgraph of four vertices without an edge.
Recall that any induced subgraph of three vertices must contain a cycle. Referring to Figure~\ref{fig:4-cycle}, there must be a directed cycle in $\{1,2,3\}$. Since there is no edge, without loss of generality, let the cycle by $1 \rightarrow 2 \rightarrow 3 \rightarrow 1$. Again, as there cannot be any edge, the cycle in $\{2,3,4\}$ must be $2 \rightarrow 3 \rightarrow 4 \rightarrow 2$. Now, we see that $\{1,3,4\}$ cannot contain any cycle without an edge (contradiction). We would have obtained the same result had we started by choosing the cycle in $\{1,2,3\}$ to be $1 \rightarrow 3 \rightarrow 2 \rightarrow 1$.
\end{IEEEproof}

\subsection{Basic Ideas}

We will prove Proposition~\ref{proposition} using the following ideas: For each category, we will show that any $G$ must contain some special arc-deleted subgraphs, say $G_\text{sub}$. For each $G_\text{sub}$, we then show that there exists a two-bit linear index code, thereby establishing $\ell^*(G_\text{sub}) \leq 2$. Since $G_\text{sub} = G^-$, from Lemma~\ref{lem:remove-add-arcs}, we must have that $\ell^*(G) \leq 2$, where the 2-bit achievability uses the same linear code for $G_\text{sub}$. As $\mais(G)=2$ is a lower bound to $\ell^*(G)$, we establish $\ell^*(G)=2$. Our approach uses combinatorics.

There are two exceptions where linear codes can only achieve three bits. For these two cases we will show that $\ell^*(G) = 3$ by calculating the chromatic number of the corresponding confusion graph. For these cases, the MAIS lower bound of two bits is loose. 

\subsection{Category 1: No undirected cycle}

We start with the first category where there cannot be any undirected cycle in $G$. We have the following subcategories:

\subsubsection{There is one or no edge} 
\begin{figure}[t]
\centering
\includegraphics[width=6cm]{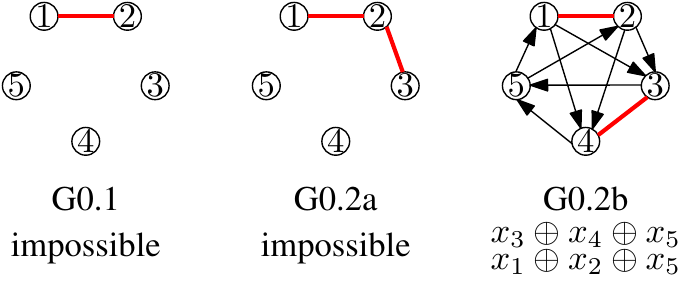}
\caption{$G_\text{sub}$ where there is one or two edges. The first two graphs are impossible for $\mais(G)=2$. For G0.2b, the two-bit index code shown here is also an index code for any $G$ (with five vertices) containing this graph.}
\label{fig:cat-1a}
\figgap
\end{figure}

If there is no edge or only one edge, we can always find an induced subgraph of four vertices with no edge.  It follows from  Lemma~\ref{lem:4-cycle} that $\mais(G)\neq 2$ (contradiction).
Figure~\ref{fig:cat-1a} shows an example where the graph G0.1 contains only one edge $1 - 2$, and the induced subgraph $\{2,3,4,5\}$ cannot contain any edge. 

Here, we use the notation G$x.y$, where $x$ is the length of the shortest undirected cycle in $G$, and $y$ is the number of edges. 

\subsubsection{There are only two edges} The two edges in $G$ can either be connected (see G0.2a in Figure~\ref{fig:cat-1a}) or disconnected (see G0.2b). We need to consider only non-isomorphic graphs, as the labeling of indices are arbitrary. 

For G0.2a, the subgraph induced by vertices $\{1,3,4,5\}$ contains no edge. By Lemma~\ref{lem:4-cycle}, this cannot happen. 

For G0.2b, since there is no edge in $\{1,4,5\}$, there must be a length-three cycle. Without loss of generality (due to symmetry), let the cycle by $1 \rightarrow 4 \rightarrow 5 \rightarrow 1$. This necessitates the cycle in $\{1,3,5\}$ to be $1 \rightarrow 3 \rightarrow 5 \rightarrow 1$. The cycles in $\{2,3,5\}$ and $\{2,4,5\}$ must also take the forms shown in the figure. For G0.2b, $\ell=2$ is achievable  by the  linear index code $[ x_3 \oplus x_4 \oplus x_5, x_1 \oplus x_2 \oplus x_5]$. 

So, any $G$ with 5 vertices, $\mais=2$, and only two edges must contain an arc-deleted subgraph isomorphic to G0.2b. By Lemma~\ref{lem:remove-add-arcs}, $\ell(G)=2$ is achievable, and hence $\ell^*(G)=2$.

\subsubsection{There are only three edges} 
\begin{figure}[t]
\centering
\includegraphics[width=6cm]{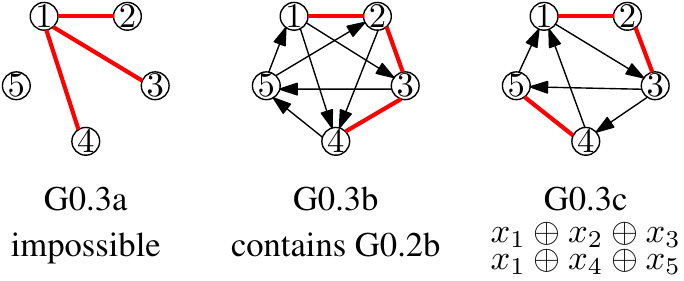}
\caption{$G_\text{sub}$ where there three edges and no undirected cycle. The first graph is impossible for $\mais(G)=2$, and there exists two-bit linear codes for the second and the third graphs.}
\label{fig:cat-1b}
\figgap
\end{figure}

Without any undirected cycle, three edges can form only three non-isomorphic configurations as depicted in Figure~\ref{fig:cat-1b}.

If the three edges form a star, we have G0.3a. By Lemma~\ref{lem:4-cycle}, it is impossible as the subgraph $\{2,3,4,5\}$ has no edge.

If the three edges form a path, we have G0.3b. The vertex set $\{1,4,5\}$ must contain a cycle. Without loss of generality (due to symmetry), let it be $1 \rightarrow 4 \rightarrow 5 \rightarrow 1$. The rest of the cycles for subgraphs with three vertices are then fixed. Since G0.3b contains G0.2b as an arc-deleted subgraph, using Lemma~\ref{lem:remove-add-arcs}, the two-bit index code for G0.2b also an index code for G0.3b.

If one of the three edges is disjoint from the other two, we have G0.3c. By symmetry and adding arcs to form cycles in $\{1,3,5\}$ and $\{1,3,4\}$, we have the configuration in the figure.  For G0.3c, $[x_1 \oplus x_2 \oplus x_3, x_1 \oplus x_4 \oplus x_5]$ is an index code.

\subsubsection{There are only four edges} 
\begin{figure}[t]
\centering
\includegraphics[width=7.5cm]{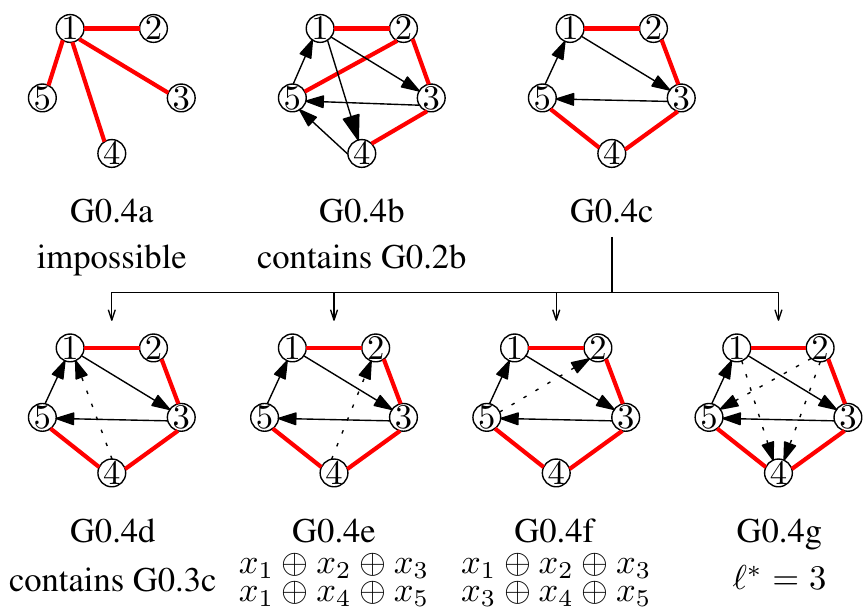}
\caption{$G_\text{sub}$ where there four edges and no undirected cycle.}
\label{fig:cat-1c}
\figgap
\end{figure}

Without any undirected cycle, four edges can form only three non-isomorphic configurations G0.4a, G0.4b, or G0.4c in Figure~\ref{fig:cat-1c}.

If the four edges form a star, i.e., G0.4a, it is an impossible subgraph as $\{2,3,4,5\}$ does not contain any edge.

For configuration G0.4b, the vertex set $\{1,4,5\}$ must contain a length-3 cycle. By symmetric let an arc in the cycle by $5 \rightarrow 1$, and so the cycle is $1 \rightarrow 4 \rightarrow 5 \rightarrow 1$. With this, the cycles for $\{1,3,5\}$ is also fixed. We see that this graph contains G0.2b as an arc-deleted subgraph, and hence the two-bit linear code for G0.2b is also an index code for G0.4b.

If the four edges form a path, we have G0.4c where we have chosen the cycle in $\{1,3,5\}$ to be (without loss of generality) $1 \rightarrow 3 \rightarrow 5 \rightarrow 1$. For G0.4c, it turns out that the best linear codes is three bits long. However,  two bits are achievable on some graphs with $|V(G)|=5$ and $\mais(G)=2$ that contain G0.4c as an arc-deleted subgraph.  So, we need to further categorize all $G$ that contain G0.4c. Since the positions of edges in G0.4c are fixed, and we can only add arcs. The only positions to add arcs are between the pairs $(1,4)$, $(2,4)$, and $(2,5)$, and we can only add at most one arc in each pair (adding arcs in both directions forms an edge). So, any $G$ in this category must satisfy either of the following:
\begin{itemize}
\item $G$ contains G0.4d, G0.,3e, and/or G0.4f as an arc-deleted subgraph, i.e., there exists an arc $4 \rightarrow 1$, $4 \rightarrow 2$, or $5 \rightarrow 2$ (from a larger index to a smaller index). The additional arcs are drawn in dotted lines in Figure~\ref{fig:cat-1c}.
\item $G$ is an arc-deleted subgraph of G0.4g (each additional arc, if present, is from a smaller index to a larger index).
\end{itemize}

A two-bit linear code exists for each of G0.4d, -e, or -f. 

For G0.4g, we calculate the chromatic number of its confusion graph to get $\ell^*(\text{G0.4g}) = 3$. It follows from Lemma~\ref{lem:remove-add-arcs} that for any arc-deleted subgraph of G0.4g, we must have that $\ell^*(\text{G0.4g}^-) \geq 3$. Now, consider some G0.4g$^-$ where $|V(\text{G0.4g}^-)|=5$ and  $\mais(\text{G0.4g}^-)=2$. We can always remove some arc(s) from G0.4g$^-$ (thereby destroying some cycles) to obtain an arc-deleted subgraph G0.4g$^{--}$, where $\mais(\text{G0.4g}^{--}) = 3$.
Using Lemma~\ref{lem:mais-2}, we know that three bits are achievable for G0.4g$^{--}$ using some linear code. Using Lemma~\ref{lem:remove-add-arcs}, it follows that the three bits linear code is also achievable for  G0.4g$^-$, and hence $\ell^*(\text{G0.4g}^-) = 3$.

\subsubsection{There are five of more edges} This configuration is impossible as it is known to contain an undirected cycle. 

So, we have shown that if $|V(G)|=5$, $\mais(G)=2$, and $G$ contains no undirected cycle, then it must either 
\begin{itemize}
\item contain G0.2b, G0.3c, G0.4e, or G0.4f as an arc-deleted subgraph, or
\item be an arc-deleted subgraph of G0.4g.
\end{itemize}
For any case, linear index codes are optimal.

\subsection{Category 2: An Undirected Cycle of Length Three}

\begin{figure}[t]
\centering
\includegraphics[width=7.5cm]{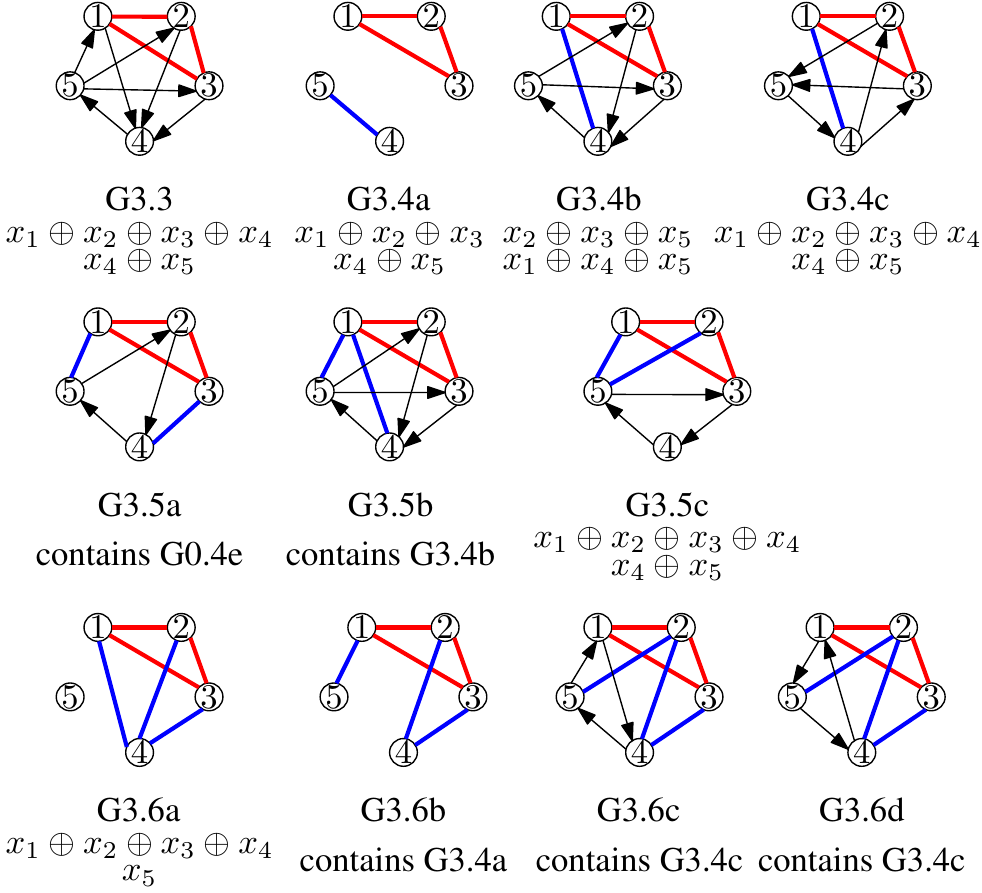}
\caption{$G_\text{sub}$ where there is a length-three undirected cycle (marked with red lines). Additional edges are marked with blue lines. Arcs are then added so that every three vertices must contain at least one cycle.}
\label{fig:cat-2}
\figgap
\end{figure}

Without loss of generality, let the undirected cycle be $1 - 2 - 3 - 1$ (depicted as red lines in Figure~\ref{fig:cat-2}). First, if there is an additional edge $4 - 5$ (denoted by G3.4a in Figure~\ref{fig:cat-2}), there exists a two-bit index code $[x_1 \oplus x_2 \oplus x_3, x_4 \oplus x_5]$. 

Otherwise (i.e., no edge between 4 and 5), any additional edge (in addition to $1 - 2 - 3 - 1$) must be between  $\{1,2,3\}$ and $\{4,5\}$. For this, we have the following categories, grouped by the number of additional edge (blue lines in Figure~\ref{fig:cat-2}):

\subsubsection{No edge between the groups $\{1,2,3\}$ and $\{4,5\}$} The only non-isomorphic graph where every three vertices contain a cycle is depicted in G3.3. For this graph, a two-bit index code is $[x_1 \oplus x_2 \oplus x_3 \oplus x_4, x_4 \oplus x_5]$.

\subsubsection{One edge between the groups} Without loss of generality, let the additional edge be $1 - 4$. Two non-isomorphic graphs with different arc positions are possible: G3.4b and G3.4c.

\subsubsection{Two edges between the groups} If the two edges connect four different vertices, we have G3.5a. If the two edges connect between the same vertex in $\{1,2,3\}$ to two different vertices in $\{4,5\}$, we have G3.5b. Otherwise, if the two edges connect between different vertices in $\{1,2,3\}$ to the same vertex in $\{4,5\}$, we have G3.5c.

\subsubsection{Three edges between the groups} The three edges can be placed in three non-isomorphic positions: (i) Between three vertices in $\{1,2,3\}$ and one vertex in $\{4,5\}$, we have G3.6a; (ii) Between three vertices in $\{1,2,3\}$ and two vertices in $\{4,5\}$, we have G3.6b; (iii) Between two vertices in $\{1,2,3\}$ and two vertices in $\{4,5\}$, we have G3.6c and G3.6d.

\subsubsection{Four or more edges between the groups} We can show that the graph will always contain G3.4a with vertex relabeling.

So, we have shown that if $|V(G)|=5$, $\mais(G)=2$, and $G$ contains an undirected cycle of length three, then there exists a linear index code of length two. 

\subsection{Category 3: An Undirected Cycle of Length Four and No Undirected Cycle of Length Three}

Next, we consider the category where there is an undirected cycle of length four; without loss of generality, let the cycle be $1 - 2 - 3 - 4 - 1$. We find the graph when there is (i) no additional edge, (ii) one additional edge, and (iii) two additional edges. Note that there cannot be three additional edge, as it will create a length-three undirected cycle. For each graph here, there exists a two-bit linear index code.

\begin{figure}[t]
\centering
\includegraphics[width=5cm]{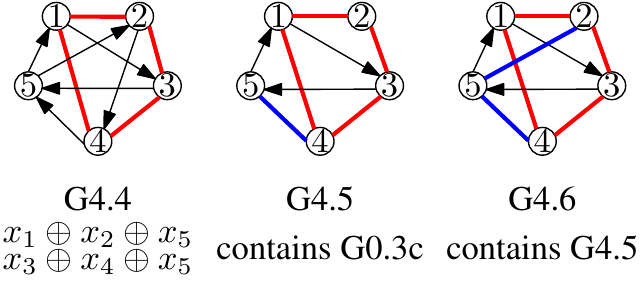}
\caption{$G_\text{sub}$ where there is a length-four undirected cycle (marked with red lines) and no length-three undirected cycle. Additional edges are marked with blue lines. Arcs are then added so that every three vertices must contain at least one cycle. There are only three non-isomorphic graphs.}
\label{fig:cat-3}
\figgap
\end{figure}

\subsection{Category 4: An Undirected Cycle of Length Five and No Undirected Cycle of Length Three or Four}

\begin{figure}[t]
\centering
\includegraphics[width=6cm]{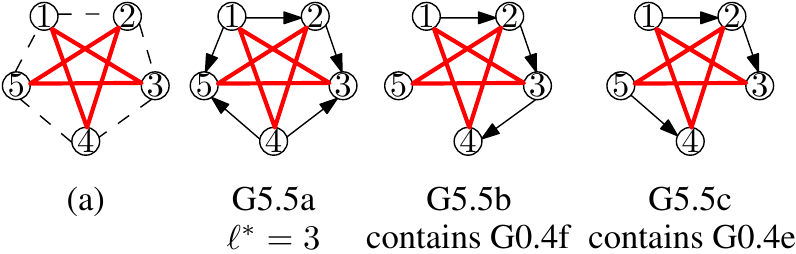}
\caption{$G_\text{sub}$ where there is a length-five undirected cycle (marked with red lines) and no length-three or -four undirected cycle.}
\label{fig:cat-4}
\figgap
\end{figure}

Without loss of generality, let the undirected cycle be $1 - 3 - 5 - 2 - 4 - 1$. With this, there cannot be any additional edge; otherwise, we get a length-three or -four cycle. Also, any additional arc must be between adjacent vertices on the ``circumference'', marked with dashed lines in Figure~\ref{fig:cat-4}(a).

As mentioned earlier in the paper, there are two configurations of $|V(G)|=5$ and  $\mais(G)=2$ where we need to calculate the chromatic number of the confusion graph to determine $\ell^*$. The graph G5.5a is one of them. For this graph, we verify (by finding the chromatic number of its confusion graph) that $\ell^*(\text{G5.5a}) = 3$. Using the same argument for G0.4g, we know that for any G5.5a$^-$ with $|V(\text{G5.5}^-)|=5$ and  $\mais(\text{G5.5}^-)=2$, we have  $\ell^*(\text{G5.5a}^-) = 3$, which is achievable by linear codes. 

We now show that for any graph in Category 4 that is not a subgraph of G5.5a, there exists a two-bit linear index code. First, if we add (i) zero, (ii) one, or (iii) two arcs to Figure~\ref{fig:cat-4}(a), we must get an isomorphic arc-deleted subgraph of G5.5a. If we add three arcs, the only graphs that are not isomorphic arc-deleted subgraphs of G5.5a are G5.5b and G5.5c. By relabeling the vertices, G5.5b contains G0.4f, and G5.5c contains G0.4e.

If we add four arcs to Figure~\ref{fig:cat-4}(a), they must form a string (i.e., a path where the direction of the arcs can be arbitrary) on the circumference (dashed lines on Figure~\ref{fig:cat-4}(a)). The only non-isomorphic combinations of length-4 strings along the circumference are: 
(i) $\rightarrow \rightarrow \rightarrow \rightarrow$, 
(ii) $\rightarrow \rightarrow \rightarrow \leftarrow$,
(iii) $\leftarrow \leftarrow \leftarrow \rightarrow$, 
(iv) $\rightarrow \rightarrow \leftarrow \leftarrow$, 
(v) $\leftarrow \leftarrow \rightarrow \rightarrow$,
(vi) $\rightarrow \rightarrow \leftarrow \rightarrow$,
(vii) $\leftarrow \leftarrow \rightarrow \leftarrow$, 
(viii) $\rightarrow \leftarrow \leftarrow \rightarrow$,
(ix) $\rightarrow \leftarrow \rightarrow \leftarrow$, and
(x) $\leftarrow \rightarrow \leftarrow \rightarrow$. 
Configurations (i)--(iii) each contain G5.5b, (iv)--(v) each contain G5.5c, (iv)--(x) each are subgraphs of G5.5a.

Lastly, we add five arcs, i.e., one arc is placed on each dash line in Figure~\ref{fig:cat-4}(a). We want to show that the graph must be G5.5a, G5.5b$^+$, or G5.5c$^+$. We can easily show that there must be a two adjacent arc in the same direction. Without loss of generality, let them be $1 \rightarrow 2 \rightarrow 3$.
For arcs between $\{3,4\}$, $\{4,5\}$, and $\{1,5\}$, if any of them does not follow the direction as that in G5.5a, we have either G5.5b$^+$ or G5.5c$^+$.


So, we have shown that if $|V(G)|=5$, $\mais(G)=2$, and $G$ contains an undirected cycle of length five, and no undirected cycle of length three or four, then it must
\begin{itemize}
\item contain G5.5c or G5.5c as an arc-deleted subgraph, or
\item be an arc-deleted subgraph of G5.5a.
\end{itemize}
For any case, linear codes are optimal. $\hfill \blacksquare$



\end{document}